\documentclass[12pt]{article}

\usepackage[english]{babel}
\usepackage[utf8]{inputenc}

\usepackage[letterpaper, margin=1in]{geometry}

\usepackage{amsmath}
\usepackage{marginnote}
\usepackage[colorinlistoftodos]{todonotes}
\usepackage{setspace}

\usepackage{cite}

\usepackage{rotating}
\usepackage{newfloat}
\usepackage{tocloft}

\usepackage{graphicx}
\usepackage{pdfpages, xfrac, pgfplots, caption, subcaption, url, ulem, bm, amssymb, animate, newfloat, ifthen}
\graphicspath{{/}{../}}

\usetikzlibrary{arrows,decorations.markings, calc}
\usepgfplotslibrary{colormaps}
\pgfplotsset{compat=1.16}
\DeclareFloatingEnvironment[fileext=frm
,name=Video]{myfloat}

\newcommand{\TR}{\textrm{TR}}
\newcommand{\TE}{\textrm{TE}}

\newcommand{\includetable}[1]{\begin{tabular} #1 \end{tabular}}

\newcommand{\marginlabel}[3]{} 
\newcommand{\figurelabel}[3]{} 
\newcommand{\remove}[1]{}

\definecolor{color1}{RGB}{0, 255, 255} 
\definecolor{color2}{RGB}{0, 0, 0} 

\newcommand{\inset}[2]{
    \def\firstview{#2}
    \def\col{#1}

    \begin{scope}
    \coordinate (origin) at (3*\shift, -4*\shift);
    \clip ($(origin)+(150:\inR*\mag)$) -- ++(150:\outR*\mag-\inR*\mag) arc (150:120:\outR*\mag) -- ++(120:-\outR*\mag+\inR*\mag) arc (120:150:\inR*\mag) -- cycle;
    \foreach \n in {0,...,11}{
        \draw[-, rrCol = \col] (origin) -- ++(0.625*12*\n+\firstview*0.625+0.625:-1*\outR*\mag);
    }
    \end{scope}
}

\newcommand{\cpOneInset}[2]{
    \def\col{#1}
    \def\firstview{#2}                

    \begin{scope}
    \coordinate (origin) at (3*\shift, 0);
    \clip ($(origin)+(210:\inR*\mag)$) -- ++(210:\outR*\mag-\inR*\mag) arc (210:240:\outR*\mag) -- ++(240:-\outR*\mag+\inR*\mag) arc (240:210:\inR*\mag) -- cycle;
    \foreach \n in {0,...,11}{
        \draw[-, rrCol = \col] (origin) -- ++(0.625*12*\n+\firstview*0.625+0.625:-1*\outR*\mag);
    }
    \end{scope}
}

\newcommand{\spokes}[6]{
    \def\col{#1}
    \def\firstview{#2}
    \def\x{#3*\shift}
    \def\y{#4*-1*\shift}
    
    \foreach \n in {0,...,11}{
        \ifthenelse{#5 = 1}{\draw[-, rrCol = \col] (\x, \y) -- ++(0.625*12*\n+\start*0.625+0.625:-1.5);}
        
        \ifthenelse{#6 = 1}{\draw[-, rrCol = \col] (3*\shift, \y) -- ++(0.625*12*\n+\start*0.625+0.625:-1.5);}
        
        \draw[-, rrCol = \col, line width=0.37 pt] (3*\shift, -4*\shift) -- ++(0.625*12*\n+\start*0.625+0.625:-1.5);
    }
    
    \ifthenelse{#5 = 1}{
        \draw[<->, color=gray] (\x-2, \y) -- ++(4, 0); 
        \draw[<->, color=gray] (\x, \y-2) -- ++(0, 4); 
    }
}

\usepackage{marginnote}

\title{BMART-Enabled Field-Map Combination of Projection-Reconstruction Phase-Cycled SSFP Cardiac Cine for Banding and Flow-Artifact Reduction}
\author{Anjali Datta, Dwight G Nishimura, and Corey A Baron \\  \\  \\  \\ 
\begin{tabular}{l} \textbf{Corresponding Author:} Anjali Datta \\ David Packard Electrical Engineering \\ 350 Jane Stanford Way, Rm. 308 \\ Stanford, CA, USA 94305-4027 \\ adatta@utexas.edu \\ (817) 269-9548 \end{tabular} \\  \\  \\  \\ }

\begin{document}

\def\offset{5}
\def\shift{4.25}

\def\rectOffset{4.5}
\def\rectSize{\rectOffset*0.732}

\def\inR{1}
\def\outR{1.5}
\def\mag{6}

\spacing{1.5}

\maketitle

{Submitted to \textit{Magnetic Resonance in Medicine}.\par}
{Approximate word count: 3200\par}

\newpage

\iftrue

\begin{abstract}
\noindent\textbf{Purpose:} To develop a method for 
banding-free bSSFP cardiac cine with substantially reduced flow artifacts. 
\newline 
\textbf{Methods:} A projection-reconstruction (PR) trajectory is proposed for a frequency-modulated cine sequence, facilitating reconstruction of three phase cycles and a field-map time series from a short, breath-held scan.  Data is also acquired during the gradient rewinders to enable generation of field maps using BMART, B$_0$ mapping using rewinding trajectories, where the rewind data forms the second TE image for calculating the field map.  A field-map-based combination method is developed which weights the phase-cycle component images to include only passband signal in the final cine images, and exclude stopband and near-band flow artifacts.
\newline
\textbf{Results:} The weights derived from the BMART-generated field maps mask out banding and near-band flow artifacts in and around the heart.  Therefore, the field-map-based phase-cycle combination, which is facilitated by the PR acquisition with BMART, results in more homogeneous blood pools and reduced hyperintense regions than root-sum-of-squares.
\newline
\textbf{Conclusion:} With the proposed techniques, using a non-Cartesian trajectory for a frequency-modulated cine sequence enables flow-artifact-reduced banding-free cardiac imaging within a short breath-hold.
\end{abstract}
\newpage

\section*{Introduction}
Balanced SSFP's high SNR, strong blood-myocardium contrast, and short acquisition time make it ideal for cardiac imaging \cite{Kramer2013}.  However, the sequence is sensitive to off-resonance -- the spectral profile of balanced SSFP (bSSFP) has periodic nulls spaced at the repetition rate (i.e., TR$^{-1}$), and images suffer from banding artifacts in regions where the local B$_0$ field strength corresponds to a null.  

In phase-cycled, or multiple-acquisition, bSSFP, images are acquired with different RF phase increments (i.e., ``phase cycles'').  The phase cycling shifts the spectral profile and therefore the bands in the individual component images, which are then combined into a null-free image \cite{Zur, Bangerter}.  Phase-cycled bSSFP is thus robust to off-resonance, but the scan time is increased by a factor of the number of phase cycles acquired, which is usually from two to four.  Various methods have been proposed to counteract this scan time increase, enabling phase-cycled bSSFP acquisitions in clinically relevant scan times \cite{Wang, Slawig, BilgicISMRM2017, Roeloffs, Datta2019}.

Another challenge to acquiring phase-cycled bSSFP in the heart is that the signal at the bands is highly sensitive to the through-plane flow rate, which results in near-band flow artifacts \cite{Markl}.  Banding falls within the blood pool in at least one of the phase-cycled component images.  Due to contributions from spins that have flowed out of the slice, flow in near-band regions results in hyperintense artifacts \cite{Markl}, which are mitigated but still present even if partial dephasing is used \cite{Datta}.  Therefore, standard phase-cycle combination methods which effectively reduce banding, such as maximum-intensity projection, magnitude-mean, and root-sum-of-squares (SOS), fail to eliminate these artifacts.  As a result, many of the proposed accelerated phase-cycling methods have not been shown in the heart \cite{Slawig, BilgicISMRM2017, Roeloffs}, and those that have, suffer from residual flow artifacts \cite{Wang, Fischer, Datta2019}.  The poor performance of standard phase-cycle combination methods for cardiac applications stems from their use of both passband signal and stopband signal, and thus artifactual near-band flow signal, in the calculation of the final combined image.  However, if the precession frequency at each location is known, it could be used to include only passband signal in the combined image, and exclude stopband and near-band flow-related signal.

For a 2D projection-reconstruction (PR) trajectory, the rewinders necessary to null the gradient axes in balanced SSFP cause the k-space trajectory to trace back along the second half of the readout path (Figure \ref{fig:BMARTsampling}a).  Therefore, by also acquiring data during the rewinders, two images with distinct TEs can be acquired with minimal modifications to the sequence for non-Cartesian bSSFP acquisitions where the rewinds fully sample k-space (Figure \ref{fig:BMARTsampling}b).  This is exploited by B$_0$ mapping using rewinding trajectories (BMART) \cite{Baron} to estimate a main-field map, which could potentially be used to determine the locations of the passbands and stopbands during phase-cycle combination, without additional scans.  


\begin{figure}
    \begin{subfigure}{\linewidth}
    \centering
        \begin{tikzpicture}
            \node at (-3, 3) {(a)};
        
            \draw[<->, line width=1pt, color=gray] (-2, 0) -- (2, 0) node[right]{$k_x$};
            \draw[<->, line width=1pt, color=gray] (0, -2) -- (0, 2) node[right]{$k_y$};
            
            \foreach \n in {0,...,15}{
                \draw[-, line width=2pt, color=black] (11.25*\n:-1.5) -- (11.25*\n:1.5);
            }
              
            \def\angle{4*11.25}
            \def\nudge{0.05}  
            \def\sqrttwo{1.414}
                        
            \draw[-, line width=2pt, color=red!67!blue!75!black] (\angle:-1.5) -- (\angle:1.5);
            \draw[<->, line width=1pt, color=gray] (-2+\offset, 0) -- (2+\offset, 0) node[right]{$k_x$};
            \draw[<->, line width=1pt, color=gray] (\offset, -2) -- (\offset, 2) node[right]{$k_y$};
            
            \draw[->, line width=1.5pt, color=red] (\offset+\nudge, 0-\nudge) -- ++(180+\angle:1.5) node[midway, below] {\small1};
            \draw[->, line width=1.5pt, color=red!67!blue!75!black] (\offset-1.5/\sqrttwo-\nudge,-1.5/\sqrttwo+\nudge) -- ++(\angle:3) node[midway, above left] {\small2};
            \draw[->, line width=1.5pt, color=blue!50!black] (\offset+1.5/\sqrttwo+\nudge,+1.5/\sqrttwo-\nudge) -- ++(180+\angle:1.5) node[midway, right] {\small3};
        \end{tikzpicture}
    \end{subfigure}
    \par\bigskip
    \begin{subfigure}{\linewidth}
    \centering
        \begin{tikzpicture}
            \node[above] at (-3, 3) {(b)};
        
            \node[above] at (0, 2.25) {Normal Echo:};
            \draw[<->, line width=1pt, color=gray] (-2, 0) -- (2, 0) node[right]{$k_x$};
            \draw[<->, line width=1pt, color=gray] (0, -2) -- (0, 2) node[right]{$k_y$};
            
            \node[above] at (\offset, 2.25) {Rewind Echo:};
            \draw[<->, line width=1pt, color=gray] (-2+\offset, 0) -- (2+\offset, 0) node[right]{$k_x$};
            \draw[<->, line width=1pt, color=gray] (\offset, -2) -- (\offset, 2) node[right]{$k_y$};
            
            
            \def\nudge{0.05}
            
            \foreach \n in {0,...,7}{
                \draw[->, line width=2pt, color=red!67!blue!75!black] (22.5*\n+11.25:-1.5) -- (22.5*\n+11.25:1.5);
                \draw[<-, line width=2pt, color=blue!50!black] (\offset, 0) -- +(22.5*\n+11.25:1.5);
            }
            \foreach \n in {8,...,15}{
                    \draw[->, line width=2pt, color=red!67!blue!75!black] (22.5*\n:-1.5) -- (22.5*\n:1.5);
                    \draw[<-, line width=2pt, color=blue!50!black] (\offset, 0) -- +(22.5*\n:1.5);
            }
        \end{tikzpicture}
    \end{subfigure}
    \caption[Illustration of the k-space coverage of the normal and rewind acquisitions.]{Illustration of the k-space coverage of the normal and rewind acquisitions.  (a) For the spoke highlighted in the illustration of the projection-reconstruction trajectory on the left, the prewinder (red), readout (purple), and rewinder (blue) are shown on the plot on the right.  (b) If neighboring spokes are acquired in opposite directions (i.e., the spokes span all $360^\circ$, not just $180^\circ$), the rewinds also sample from all of k-space.  Therefore, by acquiring during the rewinders, a second image with a later TE can be reconstructed from the bSSFP acquisition.  Note that, for simplicity, only a small number of spokes are shown here.}
    \label{fig:BMARTsampling}
\end{figure}
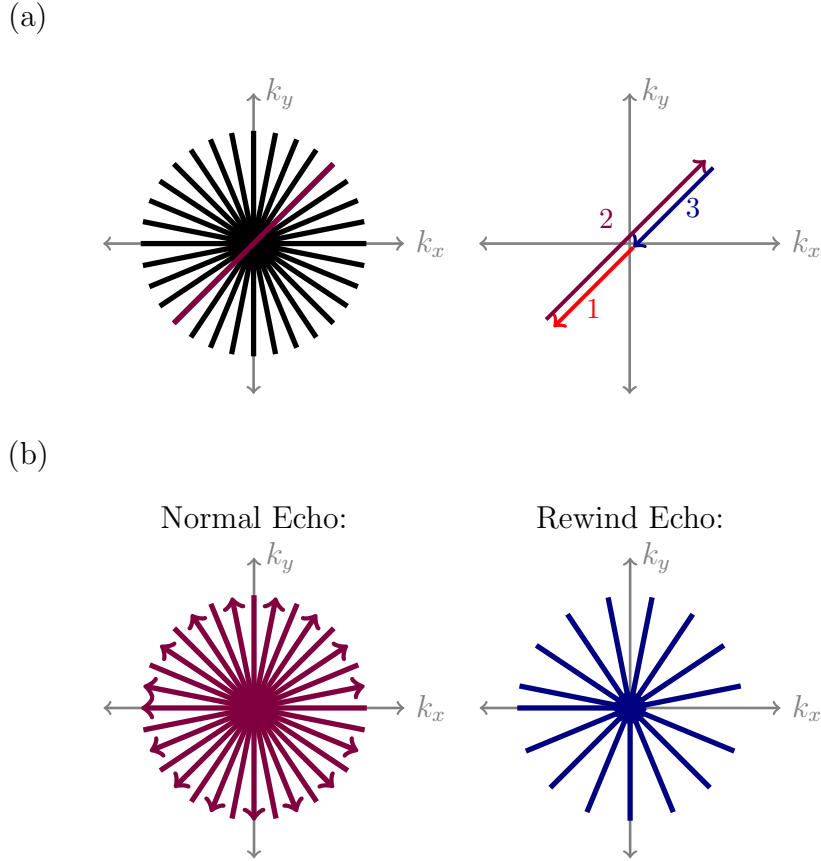


In this work, we demonstrate the feasibility of using a PR trajectory for an accelerated frequency-modulated cine sequence in conjunction with BMART to acquire three phase cycles and a field-map time series within a short breath-hold, enabling use of a field-map-based combination method when reconstructing the final cine images.  We hypothesize that, by including only passband signal, and excluding stopband and near-band flow-related signal, such a combination will result in reduced signal contributions from out-of-slice spins and a more homogeneous blood pool than root-sum-of-squares.

\section*{Methods}
\subsection*{Acquisition}

\pgfplotsset{
    colormap={inferno}{
        rgb255=(14, 9, 43)
        rgb255=(44, 11, 88)
        rgb255=(78, 13, 108)
        rgb255=(109, 24, 111)
        rgb255=(141, 36, 105)
        rgb255=(172, 48, 93)
        rgb255=(202, 64, 75)
        rgb255=(226, 87, 53)
        rgb255=(243, 117, 28)
        rgb255=(252, 151, 6)
        rgb255=(252, 189, 34)
        rgb255=(244, 228, 91)
    }
}

\tikzset{every picture/.style={line width=0.75pt}}
\tikzset{
    rrCol/.style={
        color of colormap={#1 of inferno, source range = 0:103.5},
        draw=., fill=.,
    },
}
\begin{figure}
    \centering
    \begin{tikzpicture}[scale = 0.8] 
        \foreach \nineTimesRR/\start in {0/1, 27/145, 54/294.5, 81/438.5}{
            \inset{\nineTimesRR}{\start}
            \cpOneInset{\nineTimesRR}{\start}
            \spokes{\nineTimesRR}{\start}{0}{0}{1}{1}
        }
        
        \node[below] at (2, 0) {$k_x$};
        \node[right] at (0, 2) {$k_y$};
        
        \foreach \nineTimesRR/\start in {9/149, 36/298.5, 63/442.5, 90/5}{
            \inset{\nineTimesRR}{\start}
            \cpOneInset{\nineTimesRR}{\start}
            \spokes{\nineTimesRR}{\start}{1}{0}{1}{1}
        }

        \foreach \nineTimesRR/\start in {18/290.5, 45/434.5, 72/9, 99/153}{
            \inset{\nineTimesRR}{\start}
            \cpOneInset{\nineTimesRR}{\start}
            \spokes{\nineTimesRR}{\start}{2}{0}{1}{1}
        }    
                    
        \draw[<->, color=gray] (-2+3*\shift, 0) -- ++(4, 0); 
        \draw[<->, color=gray] (3*\shift, -2) -- ++(0, 4); 
        
        \coordinate (origin) at (3*\shift, -0);
        \draw[very thick] ($(origin)+(210:\inR*\mag)$) -- ++(210:\outR*\mag-\inR*\mag) arc (210:240:\outR*\mag) -- ++(240:-\outR*\mag+\inR*\mag) arc (240:210:\inR*\mag) -- cycle;
        \draw[very thick] ($(origin)+(210:\inR)$) -- ++(210:\outR-\inR) arc (210:240:\outR) -- ++(240:-\outR+\inR) arc (240:210:\inR) -- cycle;
        \draw[dotted, very thick] (origin) -- ++(210:\inR*\mag);
        \draw[dotted, very thick] (origin) -- ++(240:\inR*\mag);

        \foreach \nineTimesRR/\start in {0/147, 27/296.5, 54/440.5, 81/3}{
            \inset{\nineTimesRR}{\start}
            \spokes{\nineTimesRR}{\start}{0}{1}{1}{0}
        }
        \foreach \nineTimesRR/\start in {9/288.5, 36/432.5, 63/7, 90/151}{
            \inset{\nineTimesRR}{\start}
            \spokes{\nineTimesRR}{\start}{1}{1}{0}{0}
        }
        \foreach \nineTimesRR/\start in {18/436.5, 45/11, 72/155, 99/292.5}{
            \inset{\nineTimesRR}{\start}
            \spokes{\nineTimesRR}{\start}{2}{1}{0}{0}
        }
        
        \foreach \nineTimesRR/\start in {0/297.5, 27/441.5, 54/4, 81/148}{
            \inset{\nineTimesRR}{\start}
            \spokes{\nineTimesRR}{\start}{0}{2}{1}{0}
        }
        \foreach \nineTimesRR/\start in {9/433.5, 36/8, 63/152, 90/289.5}{
            \inset{\nineTimesRR}{\start}
            \spokes{\nineTimesRR}{\start}{1}{2}{0}{0}
        }
        \foreach \nineTimesRR/\start in {18/12, 45/156, 72/293.5, 99/437.5}{
            \inset{\nineTimesRR}{\start}
            \spokes{\nineTimesRR}{\start}{2}{2}{0}{0}
        }
                    
        \foreach \nineTimesRR/\start in {0/439.5, 27/2, 54/146, 81/295.5}{
            \inset{\nineTimesRR}{\start}
            \spokes{\nineTimesRR}{\start}{0}{3}{1}{0}
        }
        \foreach \nineTimesRR/\start in {9/6, 36/150, 63/299.5, 90/443.5}{
            \inset{\nineTimesRR}{\start}
            \spokes{\nineTimesRR}{\start}{1}{3}{0}{0}
        }
        \foreach \nineTimesRR/\start in {18/154, 45/291.5, 72/435.5, 99/10}{
            \inset{\nineTimesRR}{\start}
            \spokes{\nineTimesRR}{\start}{2}{3}{0}{0}
        }

        \draw[<->, color=gray] (-2+3*\shift, -4*\shift) -- ++(4, 0); 
        \draw[<->, color=gray] (3*\shift, -4*\shift -2) -- ++(0, 4); 
          
        \foreach \n/\label in {0/Phase Cycle 1, 1/Phase Cycle 2, 2/Phase Cycle 3, 3/Combination}
            \node at (\n*\shift, 2.75) {\label};
            
        \foreach \n/\label in {0/Cardiac Phase 1, 1/Cardiac Phase 2, 2/Cardiac Phase 3, 3/Cardiac Phase 4} 
            \node[rotate=90] at (-2.75, -\n*\shift) {\label}; 
            
        \node at (3*\shift, -4*\shift+2.75) {Field Map};
        
        \coordinate (origin) at (3*\shift, -4*\shift);
        \draw[very thick] ($(origin)+(150:\inR*\mag)$) -- ++(150:\outR*\mag-\inR*\mag) arc (150:120:\outR*\mag) -- ++(120:-\outR*\mag+\inR*\mag) arc (120:150:\inR*\mag) -- cycle;
        \draw[very thick] ($(origin)+(150:\inR)$) -- ++(150:\outR-\inR) arc (150:120:\outR) -- ++(120:-\outR+\inR) arc (120:150:\inR) -- cycle;
        \draw[dotted, very thick] (origin) -- ++(120:\inR*\mag);
        \draw[dotted, very thick] (origin) -- ++(150:\inR*\mag);
        
        \def\scale{0.075}
        \def\xoff{1}
        \def\yoff{1}
        
        \node at (-\xoff-9*\scale, -4*\shift-\yoff+9/2*\scale) {RR};
        \foreach \RR in {1, ..., 12}
            \node at (9*\scale*\RR-9/2*\scale-\xoff, -4*\shift-\yoff+1) {\RR};
        \foreach \nineTimesRR in {0, 9, ..., 99}
            \fill[rrCol = \nineTimesRR] (\scale*\nineTimesRR-\xoff, -4*\shift-\yoff) rectangle ++(9*\scale, 9*\scale); 
    \end{tikzpicture}
    \caption[View-angle ordering for the proposed frequency-modulated PR cine sequence.]{View-angle ordering for the proposed frequency-modulated projection-reconstruction cine sequence.  Although the sequence acquires diameters of k-space, for clarity, only half of each readout is shown.  The sampling pattern is rotated every cardiac phase and between effective phase cycles.  Since different view angles are sampled for the three phase cycles, the phase-cycle combination is less undersampled than the individual phase cycles.  Since the view angles sampled during any four consecutive cardiac phases are also different, the sliding windows of complex-summed combinations used to estimate field maps are even more densely sampled.  Note that, since diameters are acquired but only radii are shown here, both $30^\circ$ insets also contain half as many lines as acquired during the normal readout -- the spokes in the opposite quadrants also traverse the highlighted region.}
    \label{fig:viewOrder}
\end{figure}
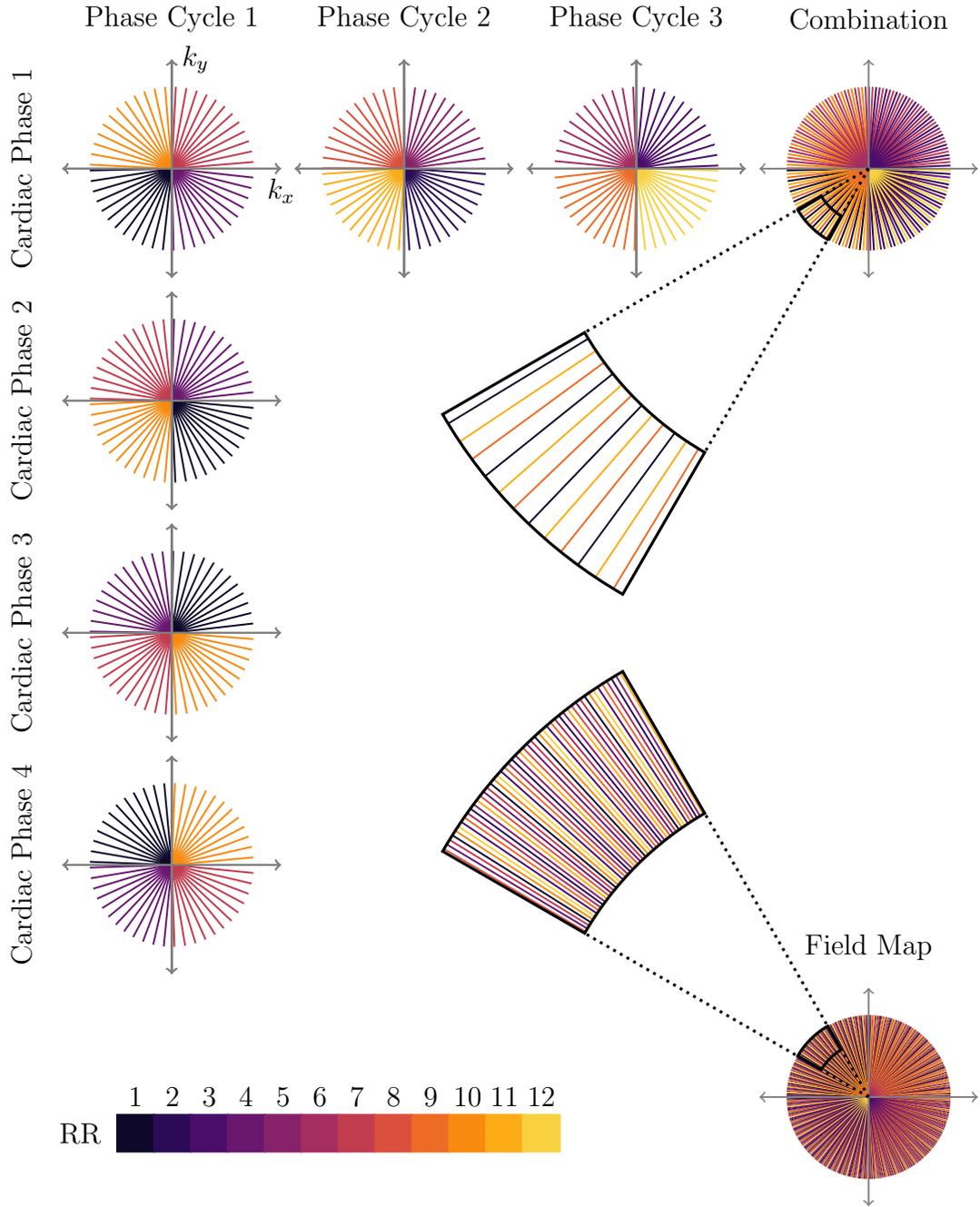

\begin{table}
    \includegraphics[width=\textwidth]{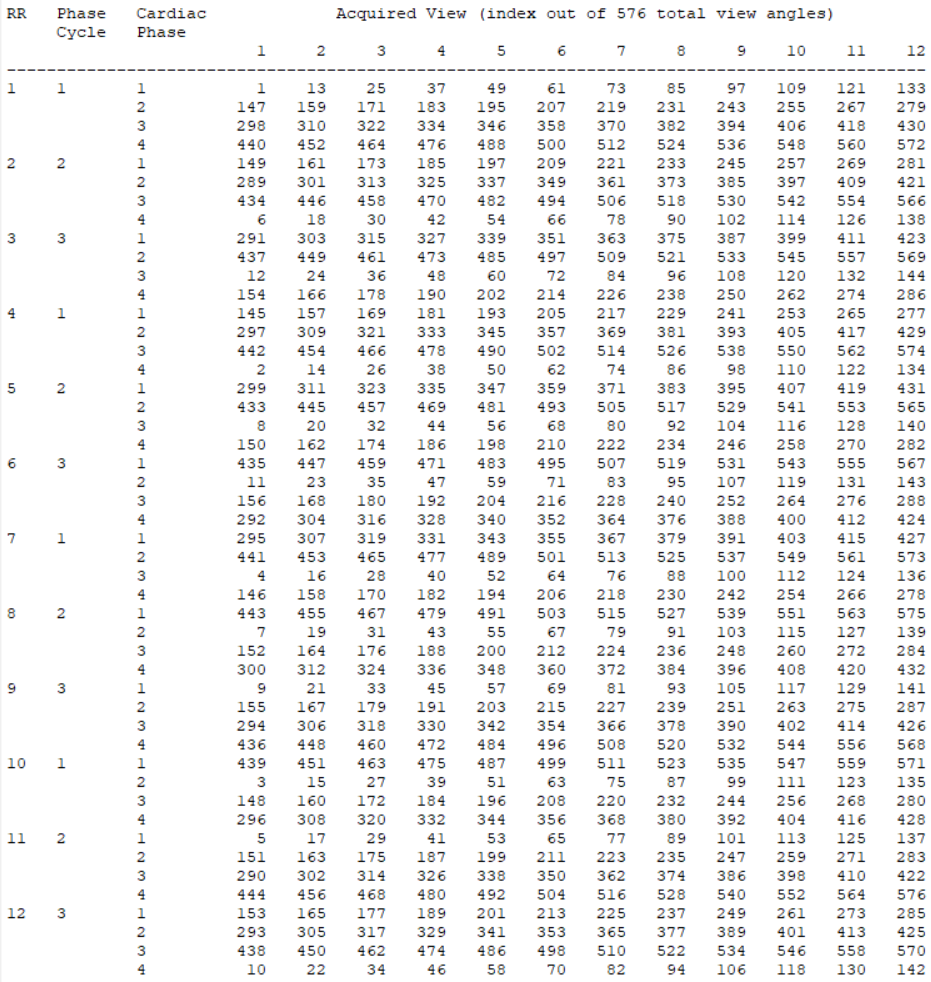}
    \caption{View angle ordering.  With 576 views, the view angle for view $v$ is $360 \cdot v/576$ degrees for $v \leq 288$ and $360 \cdot (v-0.5)/576$ degrees otherwise.}
    \label{table:view_order}
\end{table}

A PR cine sequence that acquires three phase cycles for each cardiac phase is proposed to balance banding reduction and scan time.  The scan time is thirteen heartbeats, including an initial heartbeat of discarded acquisitions for signal stabilization.  The frequency modulation scheme presented in \cite{Datta2019} is used, obviating the need for any stabilization heartbeats between phase cycles.  During the first RR interval after the discarded acquisitions, some data for phase cycle 1 is acquired.  During RRs 2 and 3, data for phase cycles 2 and 3, respectively, are acquired, before circling back to phase cycle 1 to acquire additional data in the following heartbeat  (see Figure 1a of \cite{Datta2019} for an illustration).  As a result, the acquisition of the three phase cycles is interleaved.

In the PR trajectory, the lines are rotated through all $360^\circ$ so that the rewinds cover all of k-space (Figure \ref{fig:BMARTsampling}).  An undersampled segmented acquisition with twelve view-angles per segment is used, but the sampling pattern is rotated every cardiac phase (Figure \ref{fig:viewOrder}).  For a given cardiac phase, the sampling pattern is also rotated every RR interval, so different view angles are acquired for each effective phase cycle.  Over the course of the scan, 48 k-space diameters are sampled for each effective phase-cycle of each cardiac phase, which corresponds to an undersampling ratio of 7.1.  Since different sets of spokes are acquired during different cardiac phases and effective phase cycles (Table \ref{table:view_order}), combining the cardiac phases and phase cycles generates fully sampled data for parallel-imaging calibration (similar to earlier Cartesian work \cite{Datta2019}) and field-map generation.  

The sequence was implemented at 1.5 T for use with an eight-channel cardiac coil.  It acquires a cine loop with 41 ms temporal resolution of an 8-mm-thick axial slice with 1.6 mm resolution and 36 cm field of view.  The other scan parameters are: $60^\circ$ tip angle, 1.3 ms TE, and 3.4 ms TR.  This TR is 0.18 ms longer than that of a sequence that does not acquire data on the rewinds.  Note that, unlike for the 3D acquisitions considered in \cite{Baron}, for a 2D balanced SSFP sequence, the TR is lengthened slightly since the readout rewinders can no longer overlap with the slice-select prewinder (Figure \ref{fig:pr_psd}).  The readout trapezoid uses a receiver bandwidth of $\pm125$ kHz, and samples are also acquired on the ramps.  The rewinder reaches the maximum gradient amplitude, so a bandwidth of $\pm250$ kHz is used.

Three slices of a healthy subject were imaged with informed consent and Institutional Review Board (IRB) approval.  As in \cite{Datta2019}, $30^\circ$ of partial dephasing was applied in the slice-select direction to mitigate near-band flow artifacts, and the shim was offset prior to imaging to create challenging off-resonance conditions.  Gradient delays were calibrated before reconstruction, but no other corrections, e.g., for eddy currents or gradient-amplifier nonlinearity, were applied.

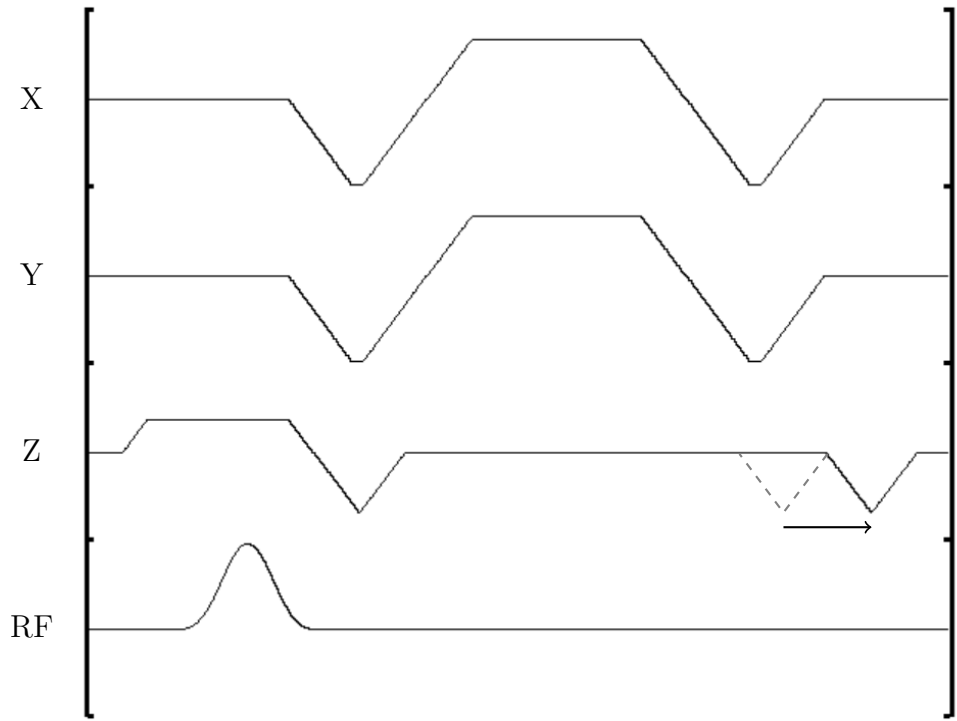
\begin{figure}
    \begin{tikzpicture}[scale=0.975]
        \node (psd) {\includegraphics[trim=30 0 0 0, clip, height=0.45\textheight, width=0.9\linewidth]{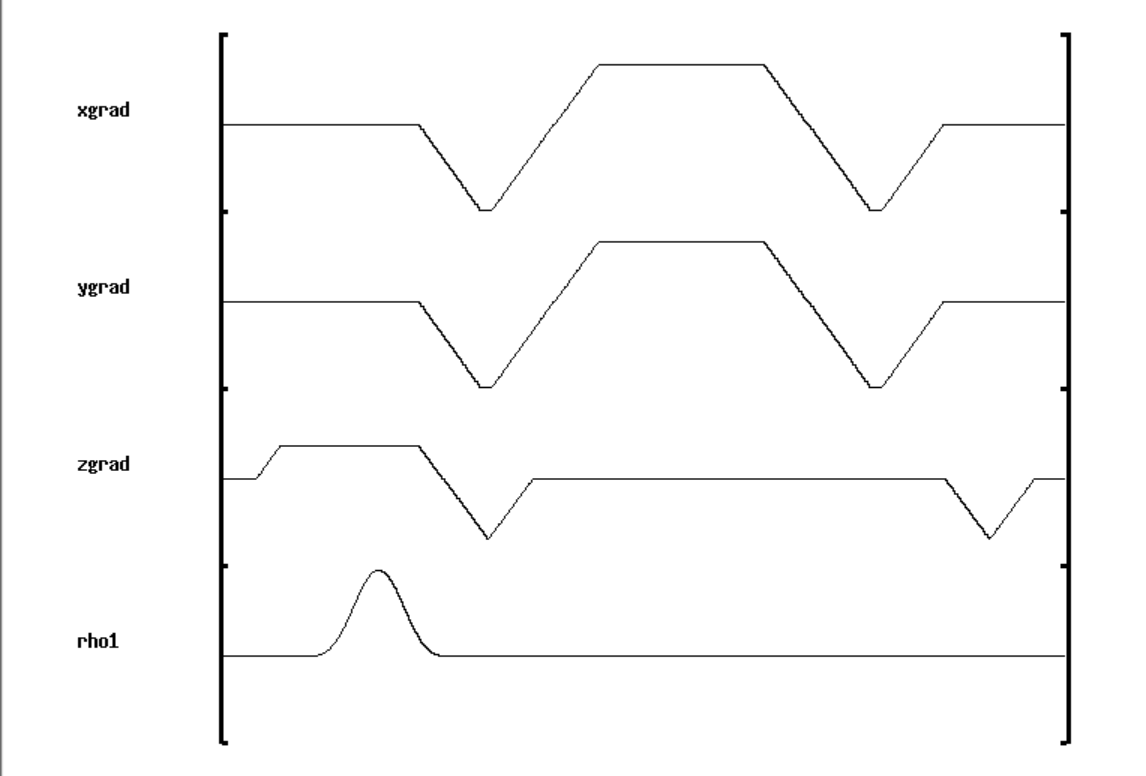}}; 
        \def \offset {-1.2}
        \draw [gray, dashed, thick] (6.05+\offset, -1.25) -- (5.45+\offset, -2.05) -- (4.85+\offset, -1.25);
        \draw [black, thick, ->] (5.45+\offset, -2.25) -- (5.45, -2.25);
        \fill[white] (-8, -5) rectangle (-6, 5);
        \node at (-6, 3.6) {X};
        \node at (-6, 1.2) {Y};
        \node at (-6, -1.2) {Z};
        \node at (-6, -3.6) {RF};
    \end{tikzpicture}
    \caption[Pulse sequence diagram for projection-reconstruction bSSFP with BMART.]{Pulse sequence diagram of projection reconstruction bSSFP acquisition that enables BMART.  Since data is being acquired on the rewinds, the readout rewinders can no longer overlap with the slice-select prewinder, so the TR is lengthened slightly.}
    \label{fig:pr_psd}
\end{figure}

\subsection*{Reconstruction}
The cine data is reconstructed using ESPIRiT with $\ell1$ spatial and temporal finite-difference regularization implemented using the SigPy Python package \cite{Ong}.  After gridding, the average of the normal k-space data (i.e., the data acquired during the readout trapezoid) is taken over all cardiac phases and phase cycles and used to estimate the receiver sensitivity maps -- the rewind data is not used for receiver-map estimation.  One slice was reconstructed with a range of different regularization parameters and then a specific parameter was empirically chosen and used to reconstruct all slices.

For the calculation of B$_0$ maps using BMART, both the normal and the rewind images are reconstructed with ESPIRiT with Tikhonov regularization using the estimated receiver sensitivity maps and an empirically determined regularization parameter.  The three effective phase cycles are complex-summed, and a sliding window four cardiac phases long is used to balance the temporal resolution of the field-map time series with the SNR of the field maps, so a total of 576 sampled lines are used for the estimation of each map.  Since neighboring spokes are acquired in opposite directions, we assume that, for the normal data, the acquisition time is TE for all of k-space (Figure \ref{fig:BMARTsampling}).  The average delay between the normal and rewind acquisition at a gridded k-space location thus depends on the acquisition times of rewind samples that contribute to that grid location.  The data is sorted into 25 bins based on this mean effective delay and the B$_0$ map is generated in MATLAB using the algorithm proposed in \cite{Baron} (see Figure 2a of that paper for an illustration).  The full width at half maximum of the Hanning window low-pass filter used to further increase the SNR of the field maps is one-third of the k-space extent.

\subsection*{Field-Map Combination}
The precession per TR, $\phi$, of the spins in each voxel is calculated from the B$_0$ map corresponding to each cardiac phase and used to weight the phase-cycled component images to emphasize passband signal and deemphasize stopband signal. If the most severe flow artifacts are localized near stopbands \cite{Markl}, such a field-map combination method should substantially reduce residual flow artifacts and contributions from out-of-slice spins in the final reconstructed image.

\begin{figure}
    \centering
    \begin{subfigure}{\linewidth}
        \subcaption{\textbf{Phase Cycle Spectral Profiles and Weighting Functions}}
        \includegraphics[width=\linewidth]{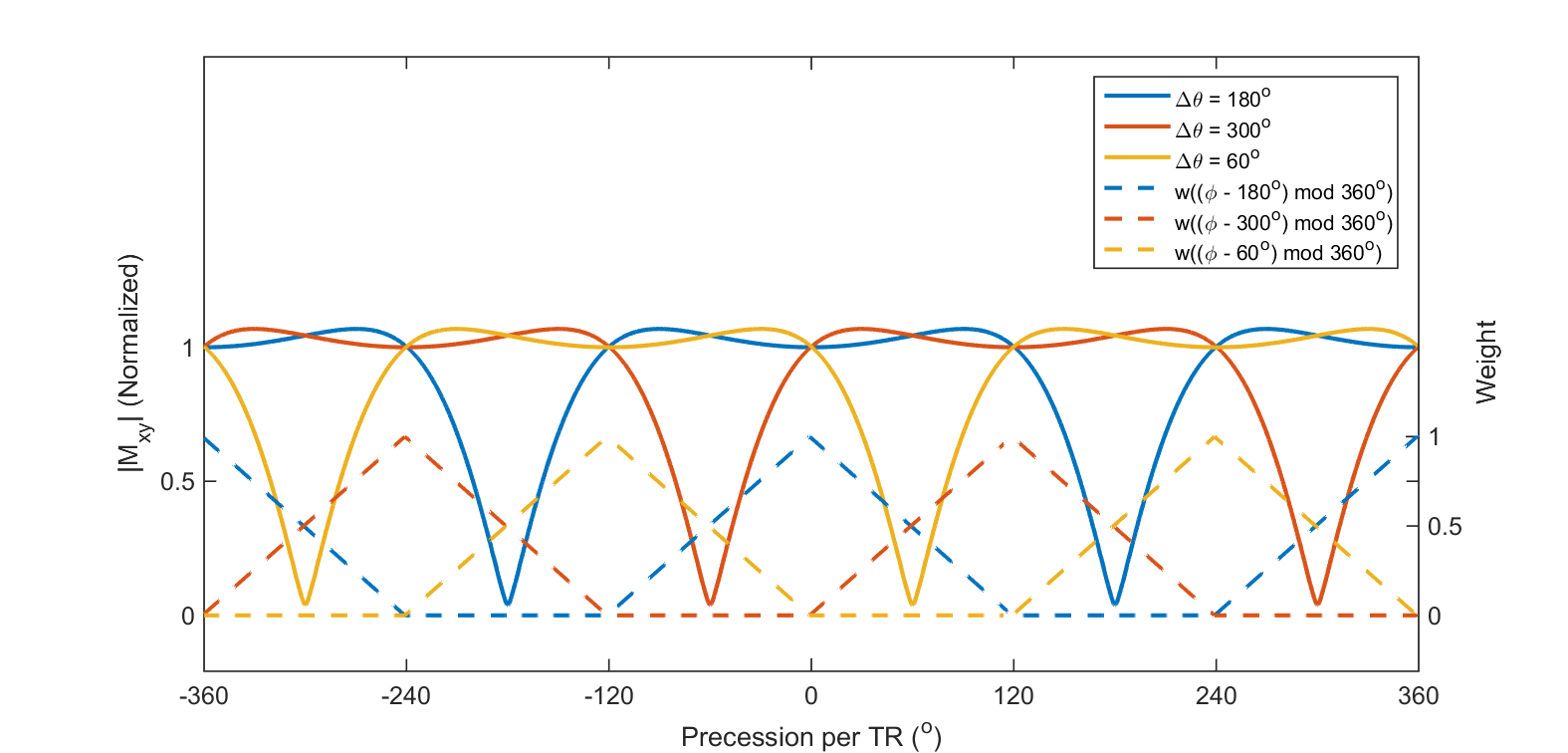}
    \end{subfigure}
    \par\bigskip\bigskip\bigskip
    \begin{subfigure}{\linewidth}
        \subcaption{\textbf{Combined Spectral Profiles}}
        \includegraphics[width=\linewidth]{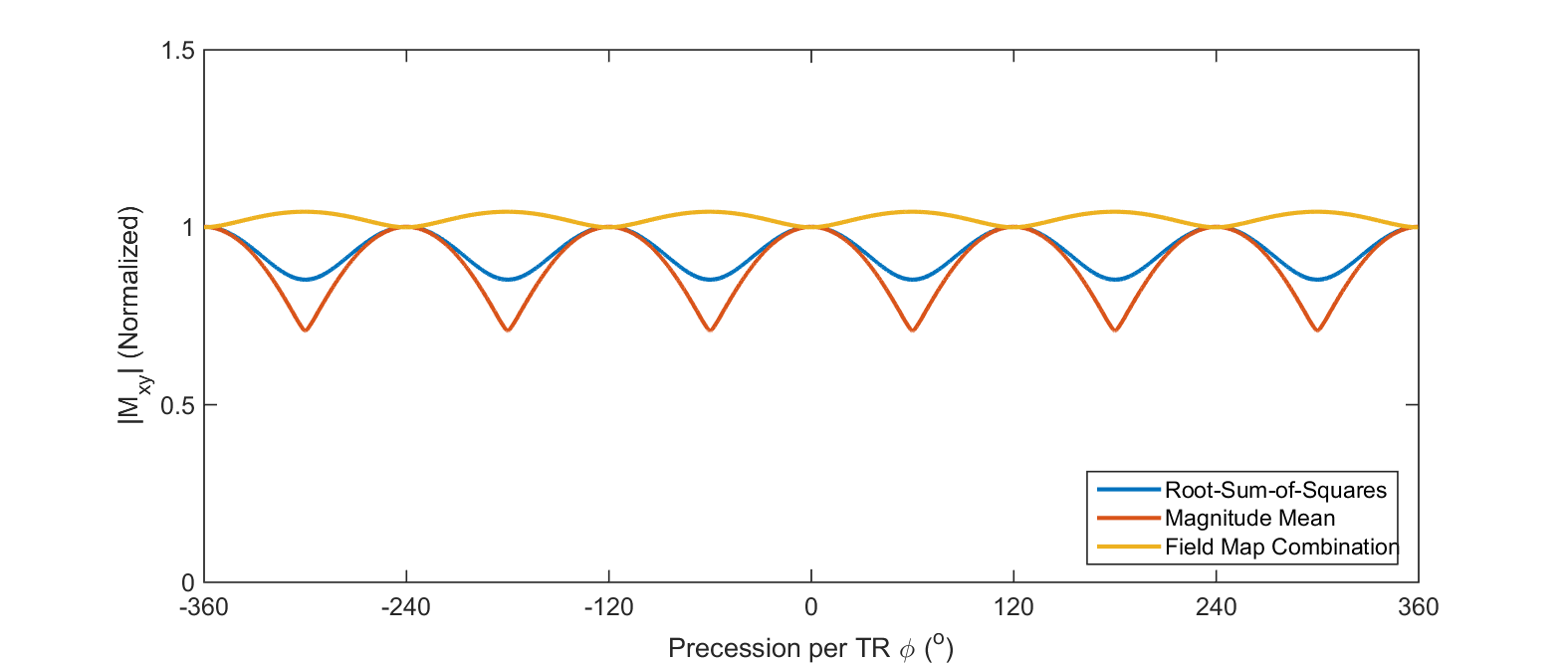}
    \end{subfigure}    
    \caption[Field-map combination of three phase cycles.]{Field-map combination of three phase cycles.  (a) The triangular weighting functions smoothly merge the passbands of the phase cycles while excluding as much of the stopbands (and associated flow artifacts) as possible.  (b) Like other combination methods, field-map combination results in a flatter spectral profile with no nulls.  The simulation parameters are $T_1 = 1000$ ms, $T_2 = 150$ ms, $\TR = 3.5$ ms, $\TE = 1.75$ ms, and tip angle $\alpha = 30^\circ$.  The individual spectral profiles are normalized such that signal at the center of any passband is 1.  Note that the relative flatness of the combined profiles depends on the $T_1$, $T_2$, and $\alpha$.} 
    \label{fig:FMC}
\end{figure}

The combined image $m_c$ is calculated as
$$m_c(x, y, t)  = \frac{\sum_{n = 1}^{N} \|m_n(x, y, t)\|w((\phi(x, y, t)-\psi_n) \mod 360^\circ)}{\sum_{n = 1}^{N} w((\phi(x, y, t)-\psi_n) \mod 360^\circ)}$$
where $x$ and $y$ are spatial coordinates, $t$ is the cardiac phase, $N$ is the number of phase cycles, $m_n$ is the nth component image, $\psi_n$ is the average phase increment for the nth effective phase cycle during that cardiac phase, and $w$ is a weighting function.  We heuristically used a triangular weighting function $w(\phi)=tri((\phi-180^\circ)/120^\circ)$ to smoothly merge the passbands of the three phase cycles while avoiding as much of the stopbands and flow artifacts as possible (Figure \ref{fig:FMC}).  Note that, over the cardiac cycle, in addition to the temporal changes in the field map, the phase cycling shifts due to the frequency modulation scheme.  Therefore, the weighting of the three images at any given location changes with the cardiac phase. In regions where the mean amplitude over the cardiac cycle of the SOS phase-cycle combination is less than 5\% of the maximum amplitude in the image, magnitude mean is used for phase-cycle combination instead of relying on the noisy field-map estimate to combine the effective phase cycles.  This is also done where the SOS amplitude is more than 50\% of the maximum, since bright voxels are presumed to consist primarily of fat, where chemical shift may result in phase wrapping in the B$_0$ map.  The proposed method is compared with root-sum-of-squares.

Bloch simulations are used to evaluate the flatness of the spectral profiles of flowing spins after field-map combination.  In particular, we look to see how effective including only the passband signal is at suppressing hyperintense signal from flow at precession frequencies near the stopbands.  The spectral profiles for various constant through-plane flow rates of SSFP with $30^\circ$ of partial dephasing are determined using the simulations described in \cite{Datta}.  The flow rate is quantified as the percent of blood in the excited slice replaced by fresh magnetization each TR.  Spin replacement percentages between $6.25\%$ and $29.2\%$ are used.  The slice thickness is 8 mm.  
Other simulation parameters are unchanged: TR $= 3.5$ ms, TE $= 1.75$ ms, $T_1 = 1000$ ms,
$T_2 = 150$ ms (in-slice), $T^*_2 = 75$ ms (out-of-slice), and tip angle $\alpha = 60^\circ$.  The profiles are normalized so that the signal of stationary, on-resonant spins in bSSFP is 1.  These profiles are then shifted to correspond to $180^\circ$, $300^\circ$, and $60^\circ$ phase-cycling, and the phase-cycled spectral profiles are combined using field-map combination and SOS.  We consider approximately two-fold enhancement of passband signal due to the inflow of fresh magnetization to be desirable \cite{Lai}, whereas we consider hyper-enhancement of near-band signal to be artifactual, so the ideal plot after phase-cycle combination would be a constant plane at approximately two (i.e., the desired signal for flowing blood is approximately twice that of static blood).

\section*{Results}
Across the range of simulated flow velocities (from 14.3 cm/s to 66.7 cm/s), field-map combination effectively excludes the hyperintense flow signal near the stopbands of the simulated phase-cycled spectral profiles.  As a result, the field-map-combined profiles are substantially flatter than those combined using SOS (Figure \ref{fig:sim_profs}).

\begin{figure}
    \begin{subfigure}{\linewidth}
        \subcaption{}
        \centering
            \includegraphics[trim=120 0 760 0, clip, width=0.95\linewidth]{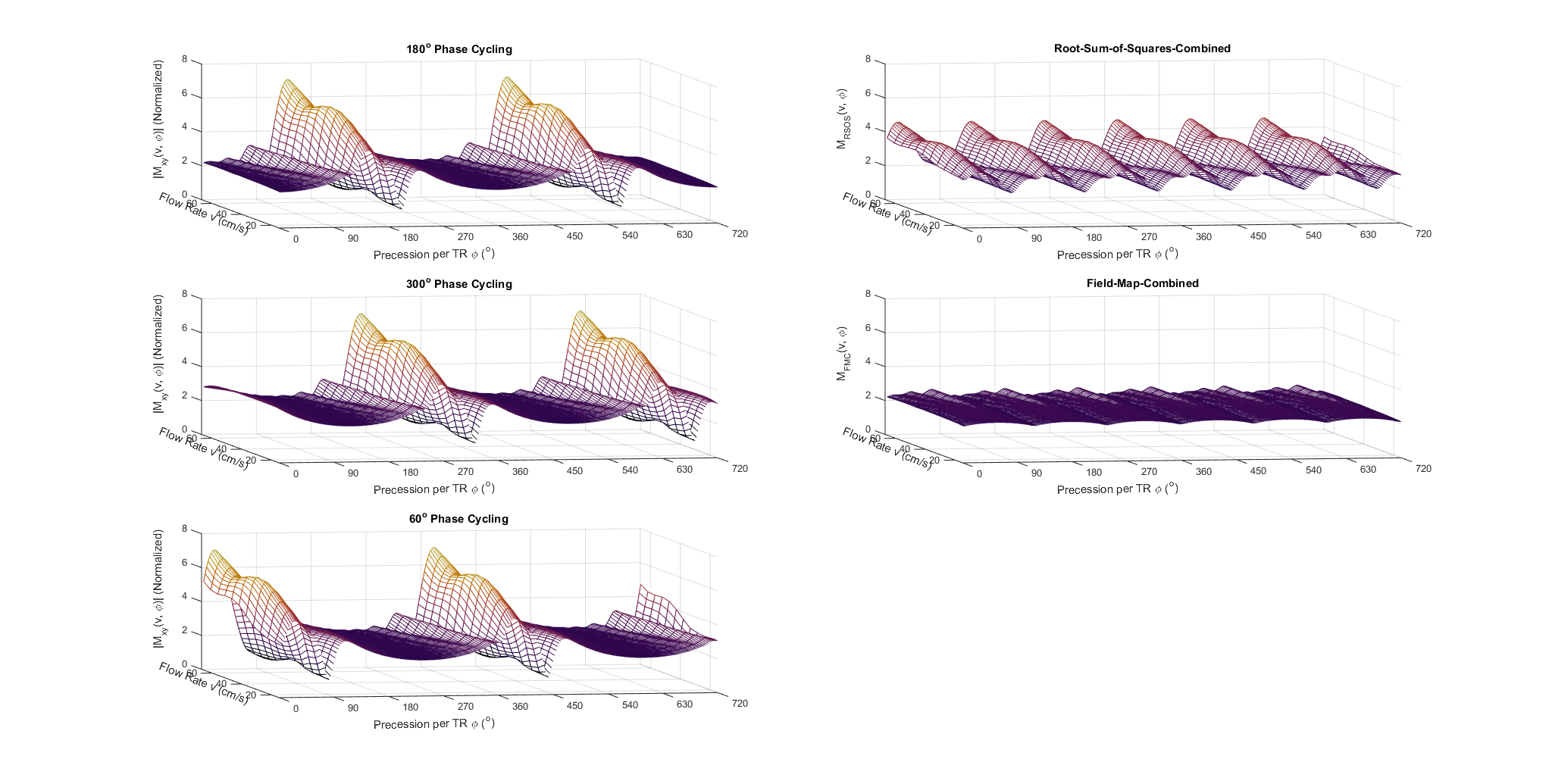}
    \end{subfigure}
    \phantomcaption
\end{figure}
    
\begin{figure} \ContinuedFloat  
    \begin{subfigure}{\linewidth}
        \subcaption{}
        \centering
            \includegraphics[trim=800 225 80 0, clip, width=0.95\linewidth]{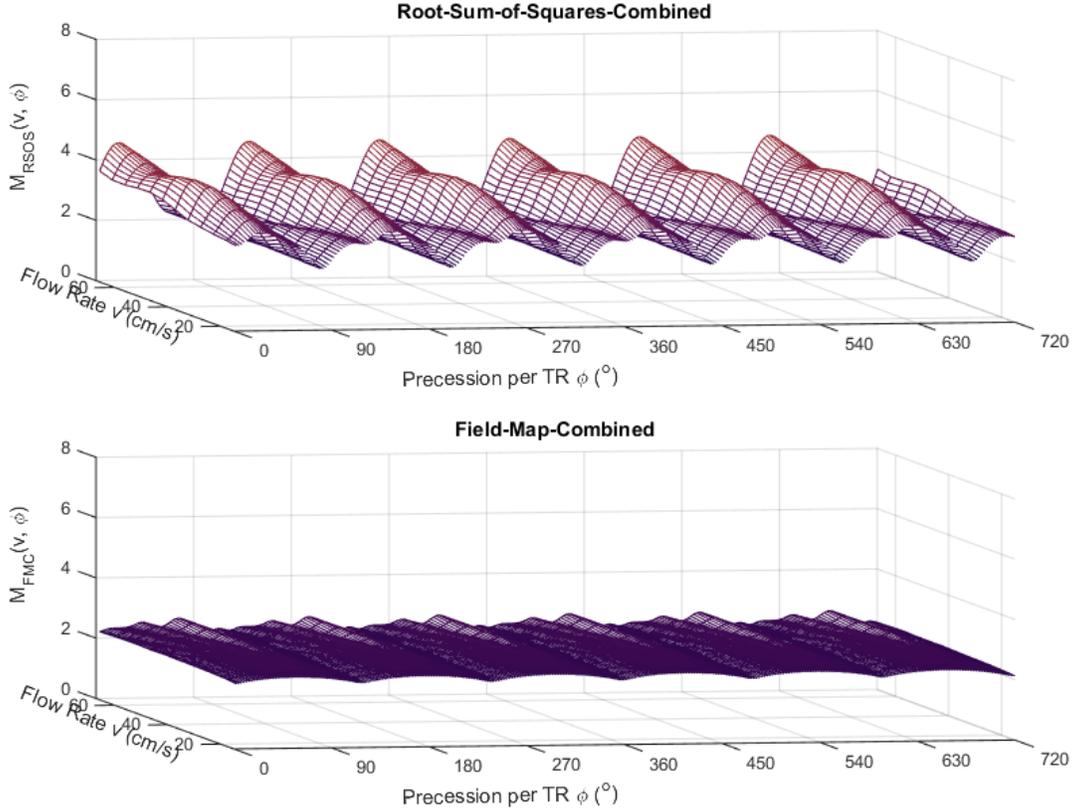}
    \end{subfigure}
    
    \caption[]{Simulated spectral profiles for various constant through-plane flow rates between 14.3 cm/s ($6.25\%$ spin replacement) and 66.7 cm/s ($29.2\%$ spin replacement).  The individual phase-cycle spectral profiles in (a) are combined using root-sum-of-squares (SOS) and the proposed field-map-based method to obtain the plots in (b).  By including only the passband signal in the final profiles and excluding the hyperintense flow signal near the bSSFP stopbands, field-map combination results in substantially flatter combined profiles than SOS.}
    \label{fig:sim_profs}
\end{figure}

In vivo, the B$_0$ maps generated by BMART appear slowly-varying over the image, as is expected, except in regions with fat or phase wrapping (Figure \ref{fig:field_map}).  In addition, the field map changes gradually over the cardiac cycle (Supporting Information Video \ref{fig:field_cine}).  The weights derived from the B$_0$ map mask out the banding and near-band flow artifacts in and around the heart in the component images (Figure \ref{fig:comp_ims}).  As a result, in all three imaged slices, the field-map combinations have more homogeneous blood pools and substantially reduced hyperintense regions than the SOS combinations (selected cardiac phases are shown in Figure \ref{fig:combined_ims}).  This holds true throughout the cardiac cycle, so the field-map-combined cine loops show less artifactual intensity variations between cardiac phases than their SOS counterparts (Supporting Information Videos \ref{fig:cine_sup}, \ref{fig:cine_mid}, and \ref{fig:cine_inf}). 

\begin{figure}
    \centering
    \textbf{BMART-Generated B$_0$ Field Maps (in Hz)}\par\medskip
    \includegraphics[trim=315 450 200 80, clip, width=0.875\linewidth]{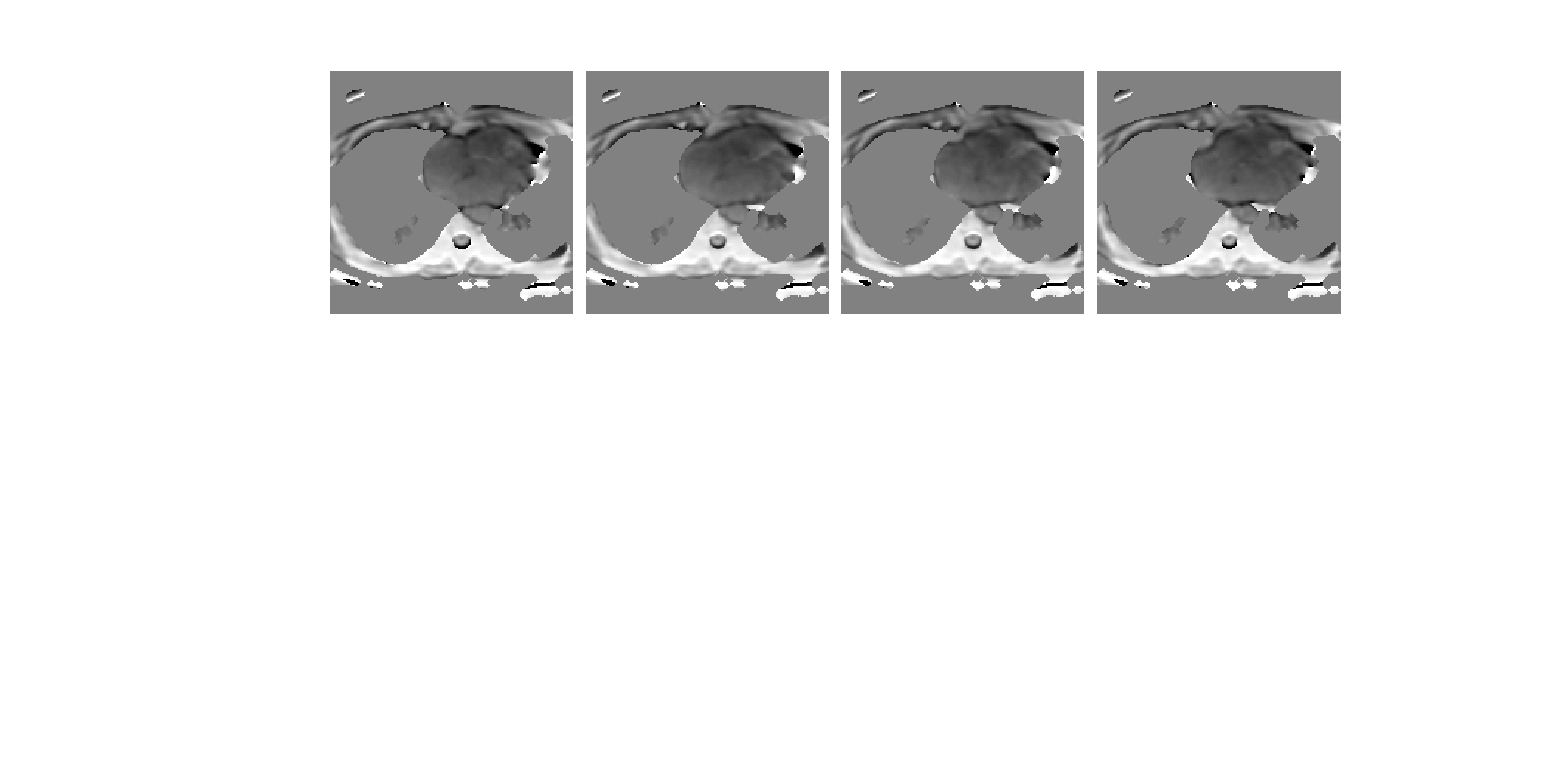} 
    \pgfplotscolorbardrawstandalone[ 
    colormap/blackwhite,
    point meta min=-400,
    point meta max=400,
    colorbar style={
        height=3.125cm, 
    }]
    \caption[BMART-generated field maps.]{BMART-generated field maps using the normal and rewind data from a sliding window of four cardiac phases surrounding each selected cardiac phase.  The cardiac phases shown here are evenly distributed through the cardiac cycle.}
    \label{fig:field_map}
\end{figure}

\begin{figure}
    \centering
    \includegraphics[width=0.95\linewidth]{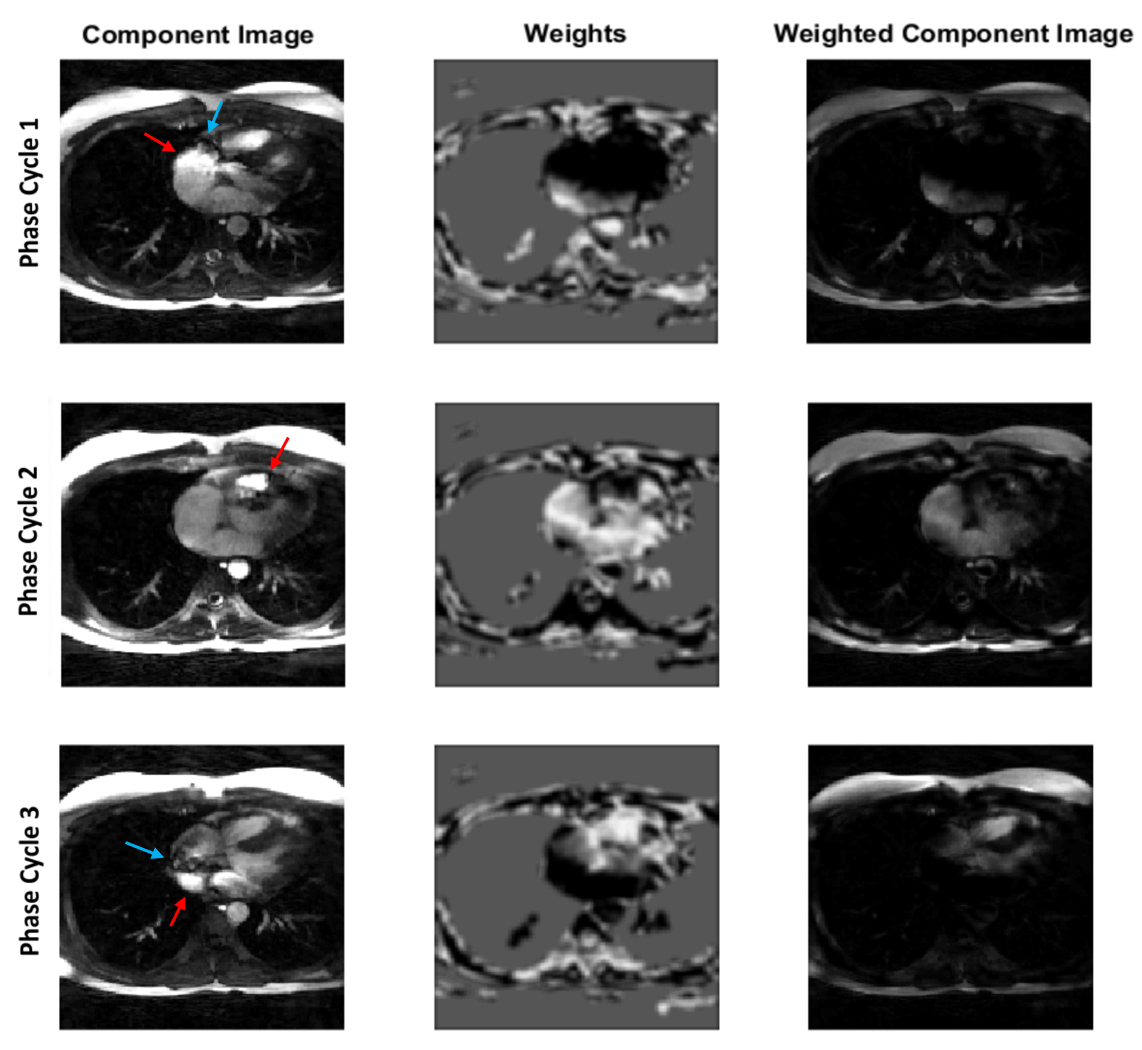}
    \caption[Weighting of component images for field-map combination.]{Weighting of component images. For all three phase-cycled component images, the weights determined from the leftmost field map in Figure \ref{fig:field_map} mask out the hyperintense flow artifacts (red arrows) near the bands (cyan arrows) in the blood pool. The field-map combination image is the sum of the weighted component images.  These component images correspond to the leftmost cardiac phase of the middle slice shown in Figure \ref{fig:combined_ims}.}
    \label{fig:comp_ims}
\end{figure}

\def\yshift{-3.5}

\begin{figure}
    \begin{subfigure}{\linewidth}
        \subcaption*{Superior Slice}
            \centering
            \begin{tikzpicture}
                \node (rsos) at (0, 0) {\includegraphics[trim=160 40 85 245, clip, width=\linewidth]{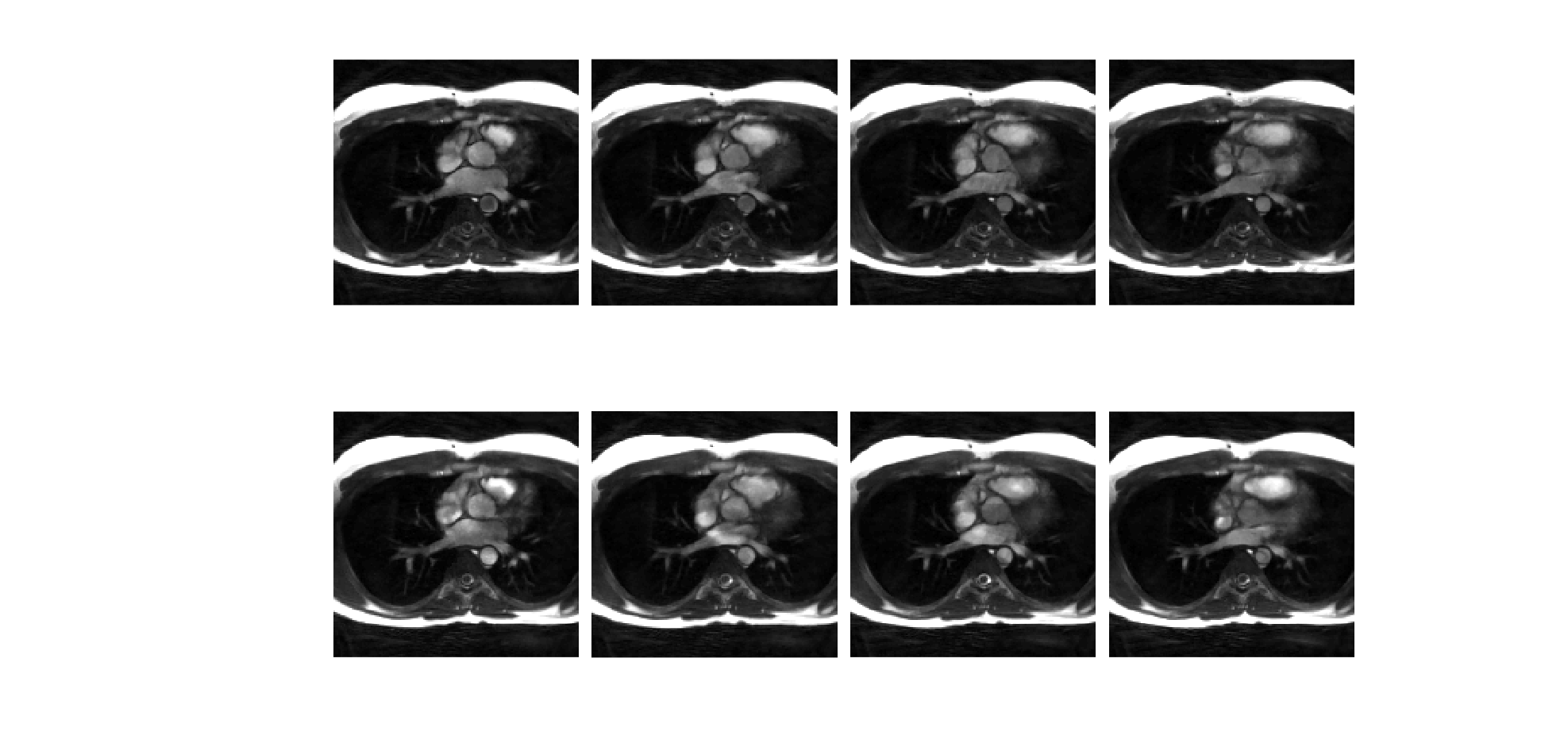}};
                \node (fmc) at (0, \yshift) {\includegraphics[trim=160 255 85 30, clip, width=\linewidth]{17408_CPs3plus4n.png}};
                \node[rotate=90] at (-8, 0) {SOS};
                \node[rotate=90] at (-8, \yshift) {FMC};
            \end{tikzpicture}
    \end{subfigure}
    \par\bigskip
    \begin{subfigure}{\linewidth}
        \subcaption*{Middle Slice}
            \centering
            \begin{tikzpicture}
                \node (rsos) at (0, 0) {\includegraphics[trim=160 40 85 245, clip, width=\linewidth]{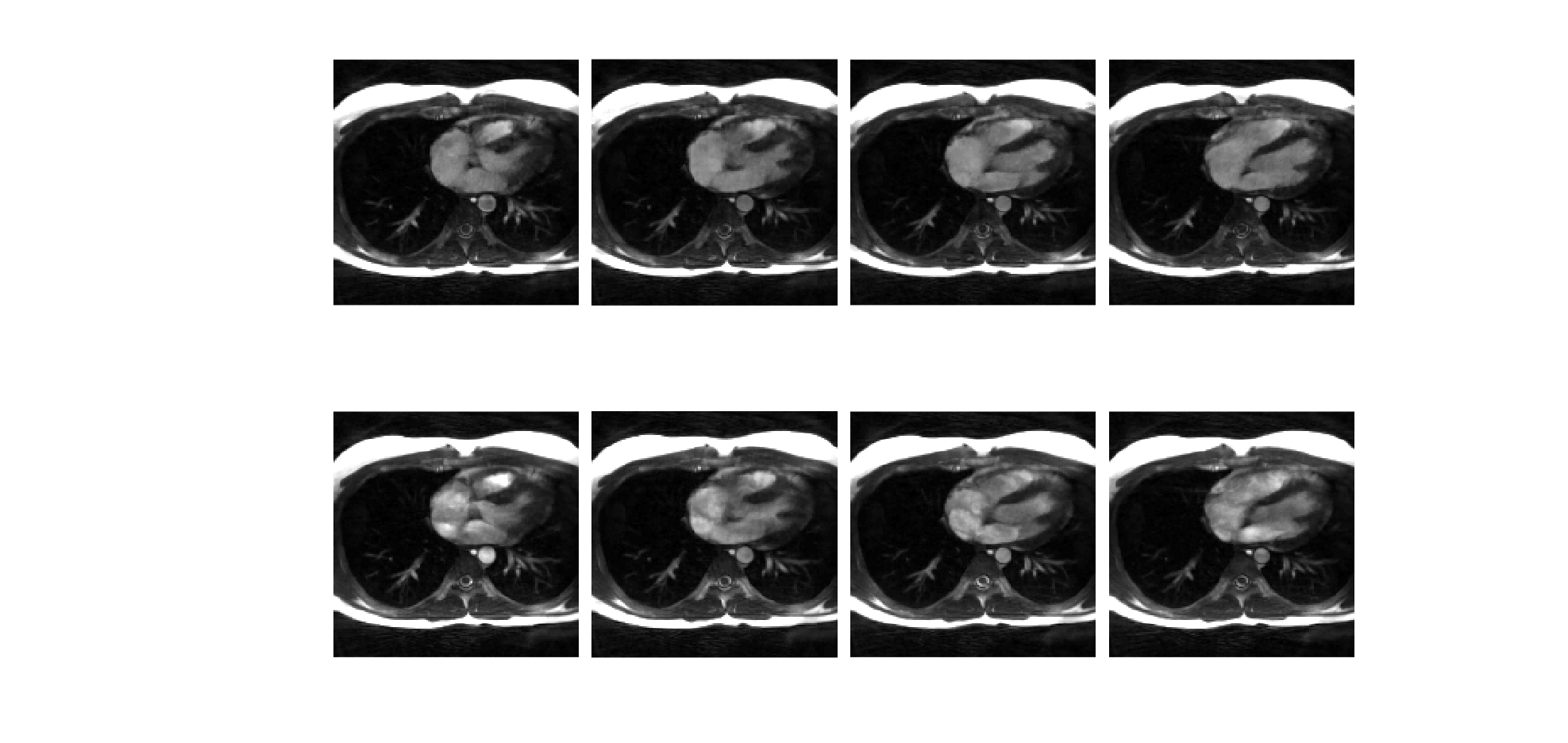}};
                \node (fmc) at (0, \yshift) {\includegraphics[trim=160 255 85 30, clip, width=\linewidth]{11776_CPs2plus4n.png}};
                \node[rotate=90] at (-8, 0) {SOS};
                \node[rotate=90] at (-8, \yshift) {FMC};
            \end{tikzpicture}
    \end{subfigure}
    \par\bigskip
        
    \phantomcaption
\end{figure}
    
\begin{figure} \ContinuedFloat   
    \begin{subfigure}{\linewidth}
        \subcaption*{Inferior Slice}
            \centering
            \begin{tikzpicture}
                \node (rsos) at (0, 0) {\includegraphics[trim=160 40 85 245, clip, width=\linewidth]{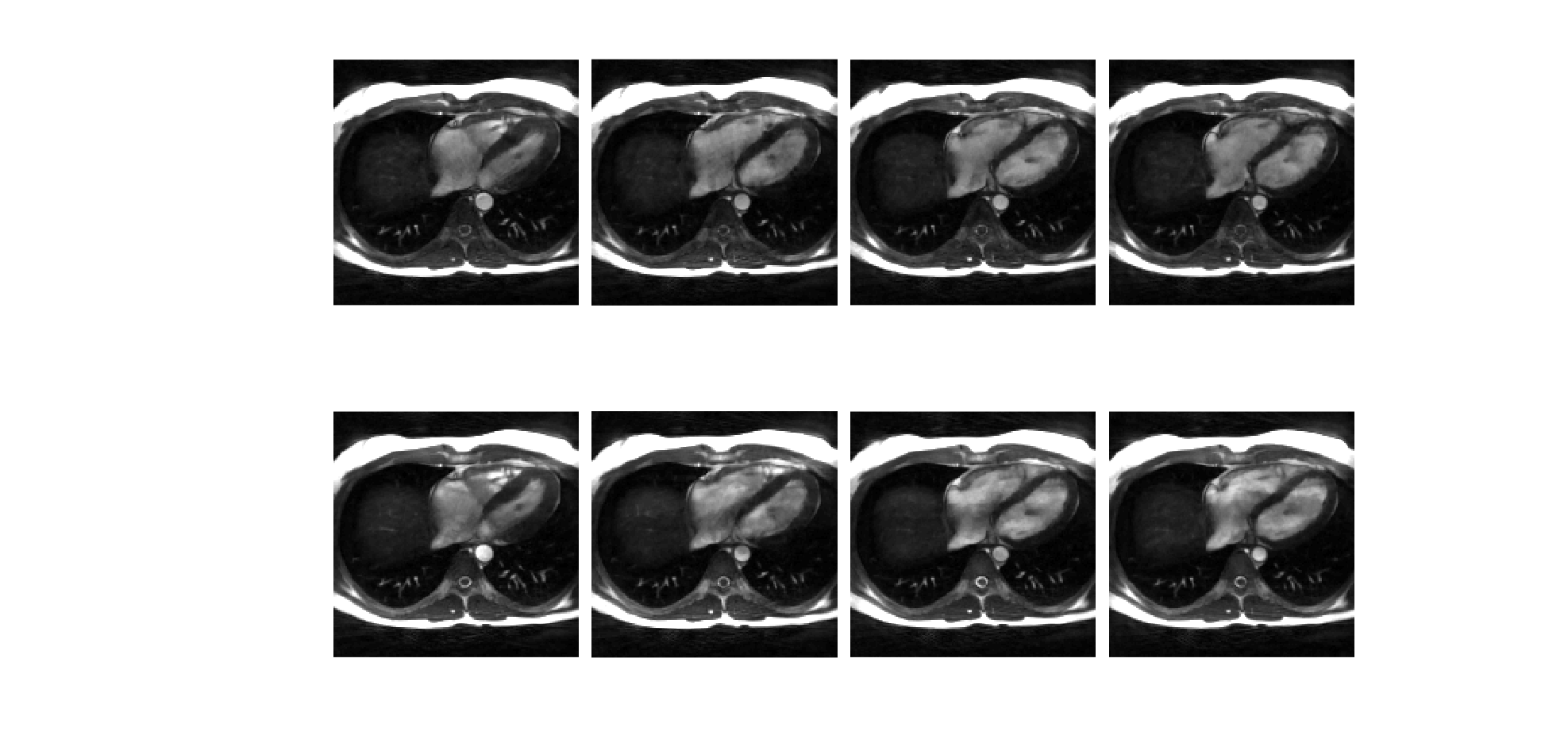}};
                \node (fmc) at (0, \yshift) {\includegraphics[trim=160 255 85 30, clip, width=\linewidth]{4096_CPs2plus4n.png}};
                \node[rotate=90] at (-8, 0) {SOS};
                \node[rotate=90] at (-8, \yshift) {FMC};
            \end{tikzpicture}
    \end{subfigure}
    \caption[Root-sum-of-squares vs. field-map combinations of three slices.] {Root-sum-of-squares and field-map combinations of cardiac phases evenly distributed through the cardiac cycle.  The field-map combinations (lower row of images for each slice, labelled FMC) have substantially reduced blood pool heterogeneity compared to root-sum-of-squares combined images (upper row for each slice, labelled SOS).}
    \label{fig:combined_ims}
\end{figure}

\section*{Discussion}
We provide a proof-of-concept that a PR trajectory can be used for a frequency-modulated cine sequence, enabling reconstruction of three phase cycles and a field-map time series from data acquired within a short breath-hold. Acquiring data on the rewinds to enable generation of field maps using BMART requires only a 5\% lengthening of the TR but facilitates field-map combination of the phase-cycle images. The resulting combined images show more homogeneous blood pool and reduced signal contributions from out-of-slice spins.

Although a Cartesian trajectory could have been used, acquiring two echoes during each TR may result in a longer scan time than PR with BMART.  On the other hand, acquiring a B$_0$ map during a separate scan, as done in \cite{DattaISMRM2017}, introduces the possibility of misregistration between the cine images and field map.  In addition, any changes in the field map over the cardiac cycle are not captured.  PR's tolerance of undersampling, motion, and flow may also be desirable in this application.

The correspondence of the weights with the banding in and near the heart suggests that the field maps generated using BMART are accurate in the region of interest.  However, the field maps suffered from phase-wrap errors in the subcutaneous fat, where the additional off-resonance from the detuned shim is also largest.  While the TE difference is 0.92 ms, note that the dynamic range of the field maps is not well-defined since the delay between the normal and rewind acquisition times decreases as $|k|$ increases; however, the range is greater than or equal to $\pm543$ Hz.  Since field-map combination only uses $\phi$ and not B$_0$ itself, selecting the time difference between the normal and rewind acquisitions at the center of k-space such that the TR is an integer multiple of it may warrant investigation; this would cause any phase-wrapped B$_0$ values to still correspond to the correct precession per TR.  It may alternatively be possible to reconstruct subspace images and linearly combine them to enable water-fat separation \cite{Roeloffs}, using the BMART-generated B$_0$-field maps to correct water-fat swaps, but this is beyond the scope of this work.  In addition, measurement of the gradient waveforms and more accurate characterization of the gradient delays may improve the accuracy of the field maps.

Note that, instead of acquiring data during the gradient rewinders, we could alternatively have acquired data during the gradient prewinders.  Acquiring data on both the prewinds and rewinds and using all three echoes to estimate the field maps may provide better robustness; however, this would double the TR increase.

When determining the field maps, instead of using the entire spoke for the normal data, using either half (either the data up to and including TE, or the data at and after TE) still results in samples from all of k-space while avoiding averaging together neighboring data points that are acquired at substantially different times.  The earlier half is an outside-in trajectory -- i.e., the same direction as the rewind data -- that results in less variation in the delay between the normal and rewind acquisitions as a function of $|k|$.  However, this causes the bin with the longest delay to no longer be at the center of k-space, which may necessitate modifications to the BMART implementation.  The latter half matches the trajectory of the rewind, with a center-out normal acquisition followed by an outside-in rewind acquisition.  However, we found that using the entire spoke resulted in field maps that appear less noisy.    

The proposed BMART-enabled phase-cycle combination should be compatible with any acquisition that results in separate phase-cycle images and that can be modified to also generate a B$_0$ map -- i.e, the frequency modulation scheme used here and partial dephasing are not necessary, and field-map combination could be trivially extended to other numbers of phase cycles (e.g., see \cite{DattaISMRM2017}).  However, partial dephasing may cause artifactual flow signal to be more localized to near the bands, which may enable field-map combination to more effectively exclude it from the final combined images.  Other spoke ordering schemes with similar properties to the one used here, such as golden-angle view ordering \cite{Winklemann}, could also be used.  
Since the frequency modulation scheme results in interleaved acquisition of the phase cycles and golden-angle view ordering covers k-space relatively evenly for any arbitrary number of views, it may be possible to reconstruct data from a smaller number of heartbeats, for example, if motion during the breath-hold is a concern.  Eddy current artifacts are a potential downside.

Although, in our experience, the flow artifact mitigation from BMART-enabled field-map combination was robust, with a marked improvement over SOS in most cardiac phases of all of the slices we acquired, some hyperintensities remained in some cardiac phases (in the right ventricle of the leftmost cardiac phase of the inferior slice in Figure \ref{fig:combined_ims}, for example).  These may have stemmed from inaccuracies in the field maps or from some flow artifacts being less localized to regions near the stopbands -- e.g., if the flow is oblique, out-of-slice spins may contribute signal to voxels away from the one they originated in.  A more sophisticated reconstruction for the B$_0$ maps that can support higher spatial and temporal resolution with adequate SNR or a more optimized weighting function in the field-map combination may address some residual artifacts, but these were not investigated here.  The basic parallel imaging reconstruction with finite-difference regularization used to reconstruct the individual phase-cycle images resulted in good quality.  However, it may be possible to better support the high acceleration by also exploiting the redundancy between the effective phase cycles, so this merits investigation.

Note that, unlike in the original paper, BMART was not used to correct off-resonance-related blurring in this work.  Because of the short readout time in the sequence used here, we did not expect noticeable blurring.  Trajectories with longer readouts, such as spirals, may enable more time-efficient acquisition, e.g., to decrease the acceleration factor or increase the number of acquired effective phase cycles.

\section*{Conclusion}
In this work, we proposed using a PR trajectory for a phase-cycled cine acquisition to facilitate estimation of a field-map time series without additional scans with BMART.  A field-map-based phase-cycle combination method was developed that utilizes the BMART-generated field maps to weight the component images to include only passband signal, and exclude stopband and near-band flow-related signal, from the final cine images.  Since the weights derived from the field maps mask out banding and near-band flow artifacts, field-map combination results in more homogeneous blood pool signal and reduced hyperintense regions compared to root-sum-of-squares.  Therefore, using the presented methods, a non-Cartesian frequency-modulated sequence can achieve flow-artifact-reduced banding-free cardiac cine imaging within a short breath-hold. 

\section*{Acknowledgements}
Thank you to GE Healthcare, NIH R01 HL127039, NSF GRFP, and the Hertz Foundation for their support.

\bibliographystyle{mrm}
\bibliography{bibliography}

\begin{thebibliography}{10}

\bibitem{Kramer2013}
Kramer~CM, Barkhausen~J, Flamm~SD, Kim~RJ, Nagel~E.
\newblock Standardized cardiovascular magnetic resonance ({CMR}) protocols 2013
  update.
\newblock J Cardiovascular Magn Reson 2013; 15:91.

\bibitem{Zur}
Zur~Y, Wood~ML, Neuringer~LJ.
\newblock Motion-insensitive, steady-state free precession imaging.
\newblock Magn Reson Med 1990; 16:444--459.

\bibitem{Bangerter}
Bangerter~NK, Hargreaves~BA, Vasanawala~SS, Pauly~JM, Gold~GE, Nishimura~DG.
\newblock Analysis of multiple-acquisition {SSFP}.
\newblock Magn Reson Med 2004; 51:1038--1047.

\bibitem{Wang}
Wang~Y, Shao~X, Martin~T, Moeller~S, Yacoub~E, Wang~DJ.
\newblock Phase-cycled simultaneous multislice balanced {SSFP} imaging with
  {CAIPIRINHA} for efficient banding reduction.
\newblock Magn Reson Med 2016; 76:1764--1774.

\bibitem{Slawig}
Slawig~A, Wech~T, Ratz~V, {Tran-Gia}~J, Neubauer~H, Bley~T, Köstler~H.
\newblock Multifrequency reconstruction for frequency-modulated {bSSFP}.
\newblock Magn Reson Med 2017; 78:2226--2235.

\bibitem{BilgicISMRM2017}
Bilgic~B, Witzel~T, Bhat~H, Wald~LL, Setsompop~K.
\newblock Joint reconstruction of phase-cycled balanced {SSFP} with constrained
  parallel imaging.
\newblock { In} Proceedings of the 25th Annual Meeting of ISMRM, Honolulu, May
  2017. p. 0441.

\bibitem{Roeloffs}
Roeloffs~V, Rosenzweig~S, Holme~HCM, Uecker~M, Frahm~J.
\newblock Frequency-modulated {SSFP} with radial sampling and subspace
  reconstruction: A time-efficient alternative to phase-cycled {bSSFP}.
\newblock Magn Reson Med 2019; 81:1566--1579.

\bibitem{Datta2019}
Datta~A, Nishimura~DG, Baron~CA.
\newblock Banding-free balanced {SSFP} cardiac cine using frequency modulation
  and phase cycle redundancy.
\newblock Magn Reson Med 2019; 82:1604--1616.

\bibitem{Markl}
Markl~M, Pelc~NJ.
\newblock On flow effects in balanced steady-state free precession imaging:
  Pictorial description, parameter dependence, and clinical implications.
\newblock J Magn Reson 2004; 20:697--705.

\bibitem{Datta}
Datta~A, Cheng~JY, Hargreaves~BA, Baron~CA, Nishimura~DG.
\newblock Mitigation of near-band balanced steady-state free precession
  through-plane flow artifacts using partial dephasing.
\newblock Magn Reson Med 2017; 79:2944--2953.

\bibitem{Fischer}
Fischer~A, Hoff~MN, Ghedin~P, Brau~AC.
\newblock Banding-artifact free {bSSFP} cine imaging using a {G}eometric
  {S}olution approach.
\newblock { In} Proceedings of the 24th Annual Meeting of ISMRM, Singapore, May
  2016. p. 1827.

\bibitem{Baron}
Baron~CA, Nishimura~DG.
\newblock B$_0$ mapping using rewinding trajectories ({BMART}).
\newblock Magn Reson Med 2017; 78:664--669.

\bibitem{Ong}
Ong~F, Lustig~M.
\newblock Sigpy: A {Python} package for high performance iterative
  reconstruction.
\newblock { In} Proceedings of the 27th Annual Meeting of ISMRM, Montreal,
  Quebec, Canada, May 2019. p. 4819.

\bibitem{Lai}
Lai~P, Cheng~JY, Vasanawala~SS, Brau~AC.
\newblock Robust free-breathing whole-heart cine {MRI} using multi-slab {3D}
  acquisition with isotropic resolution and offline reformattability.
\newblock { In} Proceedings of the 23rd Annual Meeting of ISMRM, Toronto,
  Ontario, Canada, June 2015. p. 4480.

\bibitem{DattaISMRM2017}
Datta~A, Nishimura~D.
\newblock Field map combination method for multiple-acquisition b{SSFP}.
\newblock { In} Proceedings of the 25th Annual Meeting of ISMRM, Honolulu, May
  2017. p. 0454.

\bibitem{Winklemann}
{Winkelmann}~S, {Schaeffter}~T, {Koehler}~T, {Eggers}~H, {Doessel}~O.
\newblock An optimal radial profile order based on the golden ratio for
  time-resolved {MRI}.
\newblock IEEE Trans Med Imaging 2007; 26:68--76.

\end{thebibliography}


\renewcommand{\includegraphics}[2][]{}
\renewcommand{\includetable}[1]{}

\captionsetup[figure]{labelsep=none,labelformat=empty,textformat=empty,list=yes}
\captionsetup[table]{labelsep=none,labelformat=empty,textformat=empty,list=yes}
\captionsetup[myfloat]{labelsep=none,labelformat=empty,textformat=empty,list=yes}

\newcommand{\beginsupporting}{
        \renewcommand{\tablename}{Supporting Information Table}
        \setcounter{table}{0}
        \renewcommand{\thetable}{S\arabic{table}}
        \renewcommand{\figurename}{Supporting Information Figure}
        \setcounter{figure}{0}
        \renewcommand{\thefigure}{S\arabic{figure}}
        \setcounter{myfloat}{0}
        \renewcommand{\myfloatname}{Supporting Information Video}
        \renewcommand{\themyfloat}{S\arabic{myfloat}}
     }
 
\beginsupporting


\begin{myfloat}
\centering
	\caption
	{Representive loop of the time series of BMART-generated field maps (fieldCineMid.avi).  For each cardiac phase, a sliding window of the four surrounding cardiac phases is used to estimate the field map.  That field map is then used to combine the phase cycles for the corresponding cardiac phase.  This loop is of the slice shown in Figures 5 and 6.}
    \label{fig:field_cine}
\end{myfloat}

\begin{myfloat}
\centering
	\caption
	{Root-sum-of-squares- vs. field-map-combined cine loops for the superior slice (cineSup.avi).}
    \label{fig:cine_sup}
\end{myfloat}

\begin{myfloat}
\centering
	\caption
	{Root-sum-of-squares- vs. field-map-combined cine loops for the middle slice (cineMid.avi).}
    \label{fig:cine_mid}
\end{myfloat}

\begin{myfloat}
\centering
	\caption
	{Root-sum-of-squares- vs. field-map-combined cine loops for the inferior slice (cineInf.avi).}
    \label{fig:cine_inf}
\end{myfloat}

\renewcommand{\listfigurename}{List of Supporting Information Figures}
\newcommand{\cftfiguredotsep}{\cftnodots}
\cftpagenumbersoff{figure}

\renewcommand{\listtablename}{List of Supporting Information Tables}
\newcommand{\cfttabledotsep}{\cftnodots}
\cftpagenumbersoff{table}
\renewcommand{\cfttabpresnum}{S}


\renewcommand{\listmyfloatname}{List of Supporting Information Videos}
\listofmyfloats

\renewcommand{\cftfigpresnum}{S}




\fi
\end{document}